\documentclass[conference]{IEEEtran}
\IEEEoverridecommandlockouts
\usepackage{cite}
\usepackage{amsmath,amssymb,amsfonts}
\usepackage{algorithmic}
\usepackage{graphicx}
\usepackage{adjustbox}
\usepackage{subcaption}
\usepackage{textcomp}
\usepackage[most]{tcolorbox}
\usepackage[table]{xcolor}
\usepackage{booktabs}
\usepackage{enumitem}
\usepackage{hyperref}

\usepackage{multirow}

\newcounter{obsno}
\newcommand{\observation}[1]{
    \begin{tcolorbox}[width=\linewidth, colback=lightgray!15,left=2pt,right=2pt,top=1pt,bottom=1pt]
        \textbf{{\small \MakeUppercase{Takeaway} }\refstepcounter{obsno}\Roman{obsno}:} {\small #1}
    \end{tcolorbox}
}
\def\BibTeX{{\rm B\kern-.05em{\sc i\kern-.025em b}\kern-.08em
    T\kern-.1667em\lower.7ex\hbox{E}\kern-.125emX}}
\begin{document}

\bstctlcite{bstControl}

\title{Enabling AI Deep Potentials for Ab Initio-quality Molecular Dynamics Simulations in GROMACS
\thanks{Pre-print submitted for publication.}
}

\author{
    \IEEEauthorblockN{Andong Hu, Luca Pennati, Stefano Markidis, and Ivy Peng}
    \IEEEauthorblockA{\textit{KTH Royal Institute of Technology}, Stockholm, Sweden \\
    \{andonghu, pennati, markidis, ivybopeng\}@kth.se}
}

\maketitle

\begin{abstract}
State-of-the-art AI deep potentials provide \emph{ab initio}-quality results, but at a fraction of the computational cost of first-principles quantum mechanical calculations, such as density functional theory. 
In this work, we bring AI deep potentials into GROMACS, a production-level Molecular Dynamics (MD) code, by integrating with DeePMD-kit that provides domain-specific deep learning (DL) models of interatomic potential energy and force fields.
In particular, we enable AI deep potentials inference across multiple DP model families and DL backends by coupling GROMACS Neural Network Potentials with the C++/CUDA backend in DeePMD-kit. 
We evaluate two recent large-atom-model architectures, DPA2 that is based on the attention mechanism and DPA3 that is based on GNN, in GROMACS using four \emph{ab initio}-quality protein-in-water benchmarks (1YRF, 1UBQ, 3LZM, 2PTC) on NVIDIA A100 and GH200 GPUs. Our results show that DPA2 delivers up to $4.23\times$ and $3.18\times$ higher throughput than DPA3 on A100 and GH200 GPUs, respectively. 
We also provide a characterization study to further contrast DPA2 and DPA3 in throughput, memory usage, and kernel-level execution on GPUs. 
Our findings identify kernel-launch overhead and domain-decomposed inference as the main optimization priorities for AI deep potentials in production MD simulations.
\end{abstract}


\section{Introduction}

Molecular Dynamics (MD) simulations~\cite{thompson2022lammps, abraham2015gromacs,amber} enable atomistic insights into complex systems that are otherwise inaccessible to direct experimentation. They are critical computational tools for material science, biochemistry, life science, and drug discovery. Despite MD simulations target atomic and molecular systems and their interactions and therefore quantum mechanics, the vast majority of MD simulations use classical physics, such as particle movers (i.e., a classical computational phase responsible for updating particle coordinates and velocity), and classical potential and pseudo-potential empirically fitting to quantum laws. 

Classical MD with empirical force fields, such as AMBER~\cite{amber}, CHARMM~\cite{MacKerell1998_charmm_forceField} and GROMACS~\cite{abraham2015gromacs}, scales efficiently on large-scale systems. However, they cannot reliably capture bond breaking/formation, charge transfer, many-body polarization, and related quantum-mechanical effects. In contrast, \emph{ab-initio} MD (AIMD) based on Density Functional Theory (DFT) computes interatomic forces and achieves near-quantum-mechanical fidelity. Yet, its $\mathcal{O}(N^3)$ scaling severely limits accessible system sizes and simulation timescales~\cite{He2018AIMDVariance}. 

Recent advances in machine learning for interatomic potentials, called AI Deep Potentials (DPs)~\cite{zhang_deep_2018}, provide a third path. Trained on potentials from quantum-mechanical reference data, DP models evaluate energies and forces via fast machine learning (ML) inference, delivering near-\emph{ab initio} accuracy at computational costs orders of magnitude lower than DFT. This capability enables \emph{ab initio}-quality MD for large biomolecular systems and long trajectories while remaining compatible with production MD simulation workflows. For instance, DeePMD-kit~\cite{wang_deepmd-kit1_2018} provides state-of-the-art DP models via a mature, widely used software stack for training and deploying DP models in multiple neural network architectures.

In this work, we first evaluate the opportunity and challenges of integrating AI DPs into production-level MD codes by performing a case study on GROMACS, a state-of-the-art MD code, with DeePMD-kit. We find that most inference stacks are Python-first and depend on deep-learning frameworks, whereas production MD engines are performance-critical C++/CUDA/MPI codes built around empirical force fields. Bridging the two worlds requires stable C/C++ interfaces, harmonized data layouts (e.g., neighbor lists and periodic boundaries), GPU execution pipelines, and HPC-native build systems. Another challenge comes from the lack of understanding of the computational cost of integrating DP models into end-to-end MD workflows. For instance, DPs inference throughput, scalability, and GPU utilization and memory usage in MD workflows are still insufficiently characterized, and thus performance bottlenecks are unclear.

We provide a stable, forward-compatible path to ab initio-quality forces inside production GROMACS pipelines by enabling multiple deep potential model families and ML backends within GROMACS. In particular, we provide a first-time tight integration of DeePMD-kit within GROMACS to enable GPU-accelerated AIMD simulation workflows. Our implementation extends GROMACS~\cite{abraham2015gromacs} Neural Network Potentials (NNPot) module to enable the deployment of state-of-the-art AI DP models. We evaluate this integration on four protein-in-water MD benchmarks using two recent large-atom-model (LAM) architectures, namely DPA2~\cite{zhang_dpa-2_2024} and DPA3~\cite{zhang_dpa-3_2025}, on NVIDIA A100 and GH200 GPUs. We then quantify the cost for enabling quantum-mechanical effects by comparing the \emph{ab initio}-quality simulations with the classical simulations without quantum-mechanical effects. Furthermore, we reveal the different feasibility of exploiting DPA2 and DPA3 models in production MD simulations by profiling at both the application and GPU-kernel levels to understand their computational intensity, memory footprint, and kernel launch behavior. Finally, we highlight optimization priorities for achieving scalable, high-fidelity AIMD simulations using AI Deep Potentials.

We summarize our contributions in this work as follows:
\begin{itemize}
  \item We identify the opportunities and challenges in adopting state-of-the-art AI Deep Potentials models in MD simulation workflows in GROMACS.
  \item We propose a design strategy to enable tight system integration and parallel inference in AI Deep Potentials MD simulations and an implementation in the latest GROMACS release (available at \url{https://github.com/HuXioAn/gromacs/tree/deepmd-oneModel}). 
  \item We evaluate four solvated proteins (1YRF, 1UBQ, 3LZM, 2PTC) using DPA2 and DPA3 large-atom models on NVIDIA A100 and GH200 GPUs.
  \item We provide a first-time end-to-end performance characterization of Deep AI MD simulations with quantum-mechanical effects in GROMACS and outline optimization priorities for DPA2 and DPA3 MD workflows.
\end{itemize}

\section{Background}
\label{sec:Background}
In MD simulations~\cite{thompson2022lammps,abraham2015gromacs,amber}, atom trajectories are computed by numerically integrating the Newton equations of motion, employing a Verlet or leap-frog algorithm. Forces acting on the atoms are calculated as the negative gradient of a classical potential energy function ($U(\mathbf{r})$), which is typically split into short and long-range contributions. Short-range terms include bonded contributions, such as bonds, angles, and dihedral interactions, and non-bonded van der Waals contributions (Lennard-Jones potential). Long-range electrostatic interactions are modeled with a Coulomb potential and are usually calculated using the Particle Mesh Ewald (PME) technique. The Coulomb term is divided into a short-range fast-decaying contribution and a long-range contribution. The former is computed in the real space through pairwise atom interactions, while the latter is calculated by solving the Poisson equation on a Cartesian grid in the Fourier space via 3D FFTs. All non-bonded short-range interactions are computed only among atoms that are within a cutoff distance by leveraging Verlet neighbor lists, thus, the computational cost drops from quadratic to linear. Thus, the simulation cost of a system of $N$ atoms and $N_g$ grid points for FFT scales as $\mathcal{O}(N)+\mathcal{O}(N_g\log N_g)$. 

In classical MD simulations, the functional form of $U(\mathbf{r})$ and the interaction parameters are provided by empirical force fields such as AMBER~\cite{amber} and CHARMM~\cite{MacKerell1998_charmm_forceField}, built by fitting analytic functional forms to quantum‑mechanical (QM) results and experimental data. In contrast, in \textit{ab initio} MD (AIMD) simulations, forces are computed on-the-fly from a QM treatment of the electron structure~\cite{Iftimie2005_abInitio_md}, employing the Born–Oppenheimer theory or the Car-Parrinello method. AIMD simulations provide higher accuracy and fidelity compared to the classical MD, since they can capture phenomena like bond‑breaking/forming, electronic polarization, and charge transfer. However, computational costs in AIMD scale at $\mathcal{O}(N^3)$ with the number of atoms, limiting simulations to only a few thousand atoms.

Deep neural network potentials and machine learning potentials (ML-Potentials) ~\cite{noe2020_mlp_review} offer a solution to this accuracy-cost tradeoff. ML-Potentials approximate the Born–Oppenheimer potential energy surface $U(\mathbf{r})$ by learning from large quantum‑mechanical datasets of energies and forces. Unlike empirical force fields that are based on predefined functional forms, ML-Potentials rely on flexible, non-linear function approximators that enhance the capability of capturing complex interactions and quantum effects. Several architectures and methods for encoding the atoms have been employed to develop different classes of ML-Potentials, including kernel-based methods, such as the Gaussian Approximation Potential (GAP), basis expansion approaches, deep neural networks, and graph neural networks \cite{bartok_gaussian_2010, zhang_deep_2018, zhang_dpa-3_2025}. Once trained, ML-Potentials can achieve near quantum‑mechanical accuracy with costs scaling only as $\mathcal{O}(N)$, promising for enabling large and long MD runs with near \textit{ab initio} fidelity.

\begin{figure}[bt]
    \centering
    \includegraphics[width=0.9\linewidth]{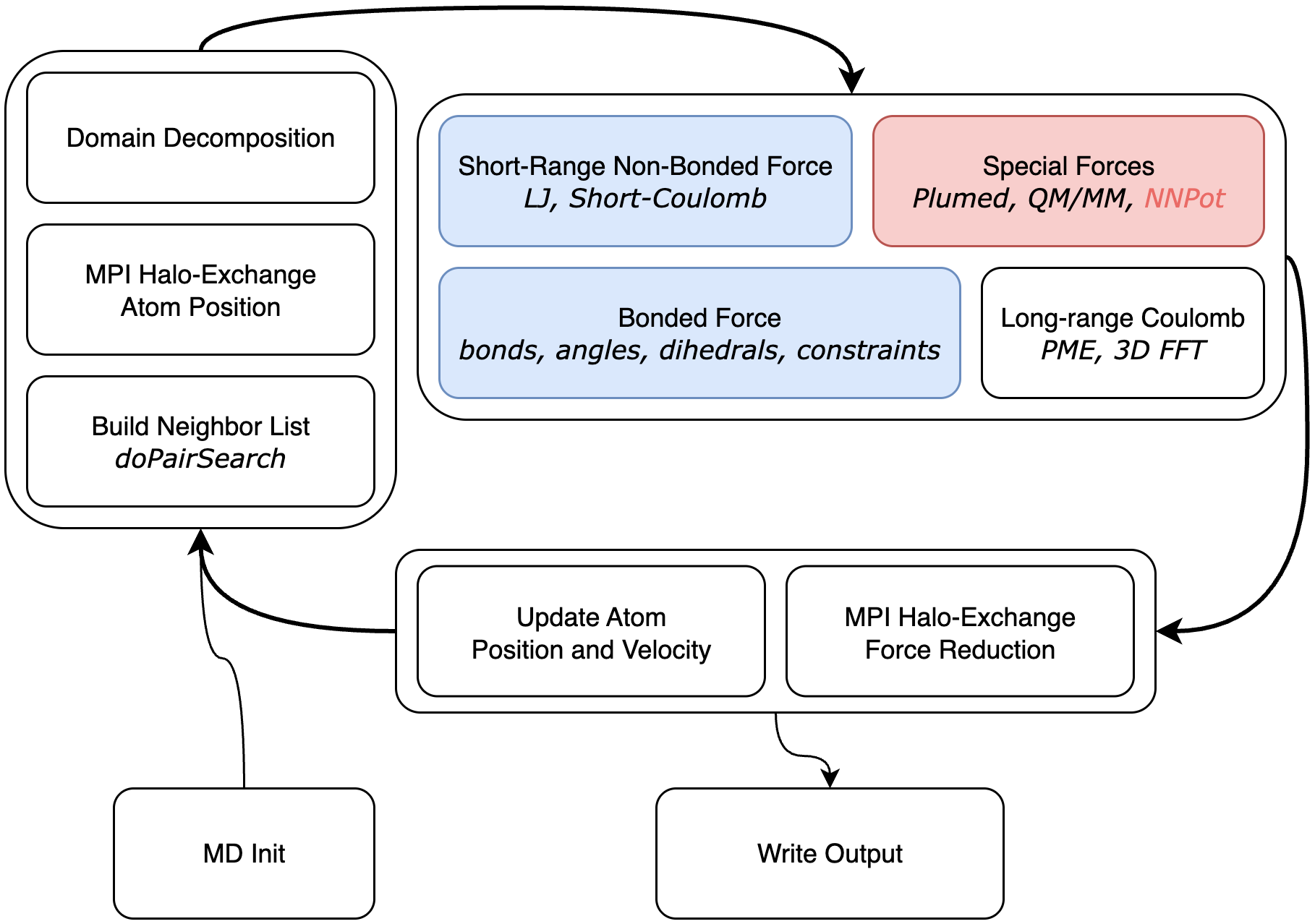}
    \caption{An overview of the GROMACS MD simulation loop.}
    \label{alg_gromacs_md_loop}
\end{figure}

\textit{GROMACS}~\cite{abraham2015gromacs} is a widely used C/C++ open-source MD engine, designed for high efficiency across a broad range of simulation scales, from laptops to supercomputers. Figure~\ref{alg_gromacs_md_loop} details the structure of the main MD simulation loop in the software. Similar to other classical MD codes, GROMACS computes the standard short and long range interactions employing empirical force fields, while additional interactions (\textit{special forces} in Figure~\ref{alg_gromacs_md_loop}) can be included in the simulation via dedicated interfaces. GROMACS introduced the ability to run MD simulations with full and hybrid neural network (NN) methods via a \texttt{\small ForceProvider} module called Neural Network Potentials. In this work, we choose to base on the latest release of GROMACS to study production MD simulation workflows. 

\textit{Deep Potential Molecular Dynamics kit (DeePMD-kit)}~\cite{Zeng2025deepmdV3} is an open-source framework designed to provide ML-Potentials models to be trained and deployed into MD engines. DeePMD-kit has a multi-backend architecture that supports NumPy, PyTorch, TensorFlow, JAX, and PaddlePaddle. The framework decouples the deep model from execution backend, improving the integration flexibility with existing MD codes. 

DeePMD-kit supports both standard deep neural network and graph neural network potentials. In this work, we choose to study two recent deep potential models from these two types, i.e., the DPA2 model using the standard deep neural network and the DPA3 model using graph neural network. 

DPA2 and DPA3 are called large-atom models (LAMs)~\cite{zhang_dpa-2_2024, zhang_dpa-3_2025}. LAMs are trained in a multi-task way, with one unified descriptor trained from a collection of datasets, and corresponding \texttt{\small fitting-nets} of each of the dataset. This scheme reduces the cost of fine-tuning the pretrained LAM for specific downstream tasks. 

DPA2, DPA3, and DPA1 belong to the DPA model family that are the latest model architectures developed by DeePMD team \cite{zhang_dpa-1_2023, zhang_dpa-2_2024, zhang_dpa-3_2025}. DPA1 and DPA2 take the advantages of the attention mechanism to capture the high-dimensional relationship between atom and its neighbors. The information of atoms and their neighbors are encoded to high-dimensional representations using a multi-layer perceptron (MLP) \texttt{\small embedded-net}, called \texttt{\small repinit} in DPA2, then updated by multiple composed self-attention layers in a residual way, called \texttt{\small repformer} in DPA2. Finally, the updated representations can be used to generate the energies on each atom in the system, through a trained MLP \texttt{\small fitting-net}. Instead of attention mechanism, DPA3 utilizes graph neural network (GNN) and Line Graph Series (LiGS) to capture and update physical representations in the system, such as bond, angle and dihedral. In this work, we use \texttt{\small DPA-2.4-7M} and \texttt{\small DPA-3.1-3M}, the latest releases of DPA2 and DPA3 by DeePMD team~\cite{zhang_dpa-3_2025}.

\section{A Deep Potential AIMD Simulation Workflow}
\label{sec:methodology}
\begin{figure}[bt]
    \centering
    \includegraphics[width=0.9\linewidth]{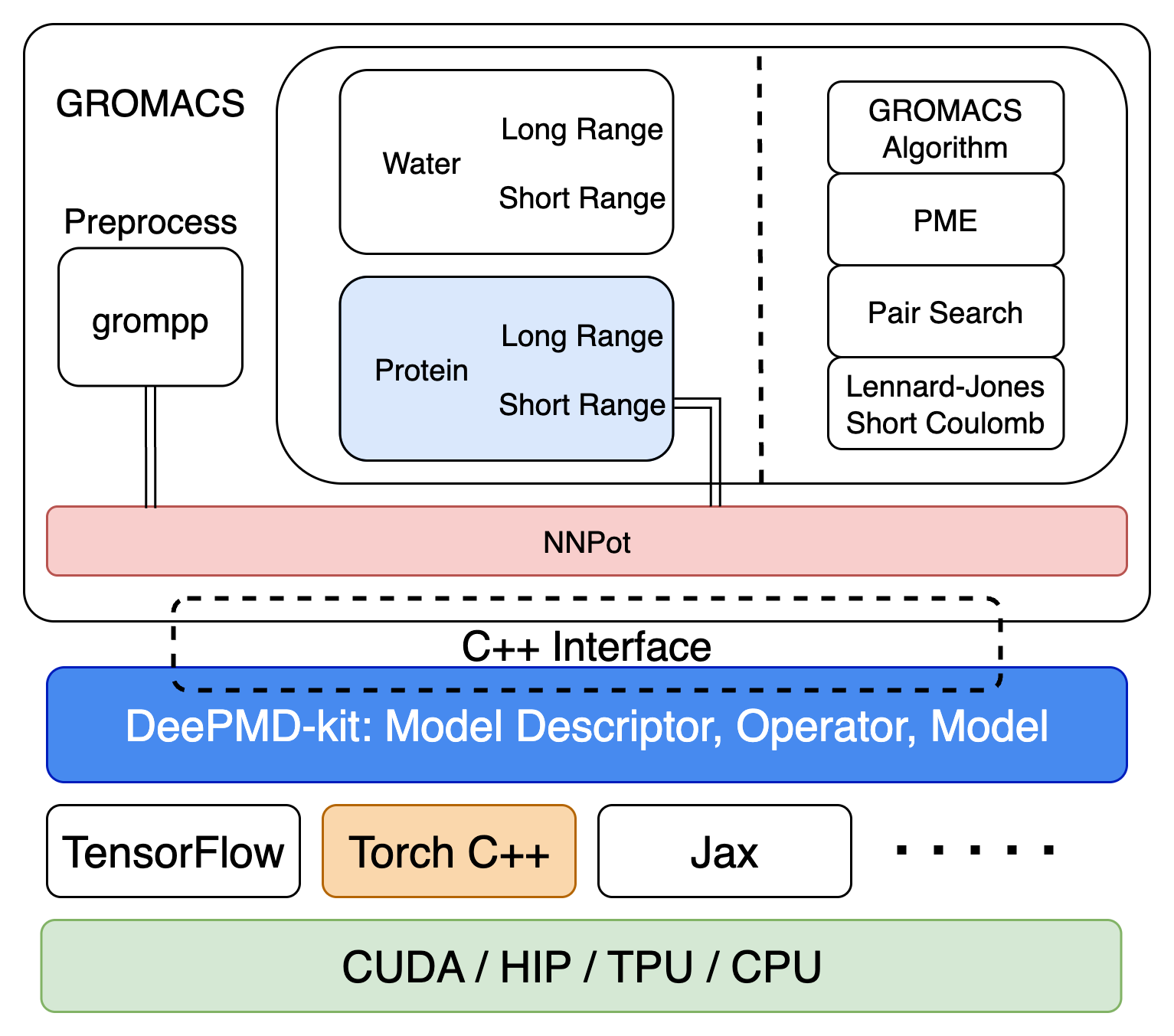}
    \caption{An overview of the GROMACS MD architecture with the DeePMD-kit integration.}
    \label{fig:integration}
\end{figure}

In this section, we introduce the main design considerations of a deep potential AIMD simulation workflow and then describe our implementation. Figure~\ref{fig:integration} presents the overall architecture of the AIMD simulation workflow coupling GROMACS and DeePMD. In this example, we take protein MD simulations as the test case to illustrate the main stages and data structures. The main data structures include water and protein and their long-range and short-range force fields. The top tier represents the main MD run inside GROMACS. The bottom tier provides the deep potential model descriptors, operators, and model training and inference. Between the two tiers, a coupling layer is highlighted in the dashed C++ interface and NNPot, which exchange the information of atoms and force fields between the two tiers.

The whole MD simulation workflow consists of multiple stages. Before the actual MD simulation, energy minimization (EM) is conducted to relax the system, ensuring a physically reasonable starting structure. It is followed by two equilibration processes, first under constant volume and temperature (NVT) conditions to stabilize the temperature, and then under constant pressure and temperature (NPT) conditions to equilibrate the system's density and pressure. Once the system is stable, MD simulation is carried out, where the actual motion of the atoms over time is simulated. The deep potential inference in DeePMD can take part in all of these stages as they need to calculate the force acting on each atom.

We design our AIMD workflow to use the deep potential models only in the final MD stage for two main considerations. First, the NVT and NPT equilibration can take tens of thousands of steps each. As they only provide a starting point for the main MD runs, no \textit{ab initio} quality is expected for this stage and thus we skip using deep potentials for this stage. Second, keeping the current preprocess allows the workflow to have a common EM and equilibration condition as the starting point for both classical simulations and deep potentials simulations.

The AIMD workflow is designed to support either full deep potential simulations or hybrid classical-deep potential simulations, in which only part of the atoms are simulated with deep potential models. In particular, the workflow can select specific atoms to be simulated via deep potential models. For instance, the water and protein atoms in a solvated protein MD simulation can be separately processed by classical methods or AIMD methods, depending on the use case scenarios. This feature balances the accuracy and performance of AIMD methods. 

In the GROMACS preprocessing stage \texttt{\small grompp}, before the final MD run stage, the NNPot module disables all the interactions inside the group \texttt{\small protein} by removing bonds and dihedrals. Additionally, it adds the target atoms to exclusion lists to avoid the atoms being included in neighbor lists, thus disabling non-bonded short-range interactions. During the MD run, NNPot forwards the atom data of the protein to DeePMD-kit for model inference to get short-range forces between protein atoms. In the example illustrated in Figure~\ref{fig:integration}, we use deep potential models for interactions inside the protein atoms but not water atoms. The interactions of water atoms, including those between the protein and water, are instead provided by the classical algorithms in GROMACS. The selection of the atom group that will be processed by the deep potential models can be easily done using the NNPot options in the \texttt{\small mdp} configuration input files.

We design our workflow to support a tight coupling between GROMACS and DeePMD-kit. GROMACS provides the NNPot module to interface with ML models, but supports only PyTorch models currently. Although DeePMD-kit provides a code patch for the GROMACS 2020.2 release by adding model inference code to the core function \texttt{\small do\_force} in GROMACS, this integration requires invasive preprocessing of the input protein. Considering the two, we integrate DeePMD-kit inference as an additional backend to the GROMACS NNPot module and enable the communication of atom information and force fields between the two.

We identify three technical challenges in enabling a tight coupling between the two in AIMD workflows. The first challenge stems from the communication nature within the DPA2 and DPA3 models. Unlike earlier deep potential models, which only take ghost atoms in an $Rcutoff$ ghost halo area, DPA2 and DPA3 need to take atoms in a $L\times Rcutoff$ ghost area according to their inner layer L. We solve it by broadening the ghost area of subdomains in GROMACS to provide needed atom information to DPA models. The second challenge arises from the asymmetric ghost topology in GROMACS. Since the classical method is linear, on the boundary of two subdomains, only one side gets the ghost atoms from the other subdomain because they just need to add the contributions from both sides on the same atom. However, deep potential models are nonlinear and require full atom information in each subdomain. We solved this with communication before inference to rebuild the correct topology and ghost atoms in each subdomain. The third challenge arises from the lack of support for parallel inference in the GROMACS Neural Network Potentials, resulting in the absence of ghost atom information. Our current design reuses the NNPot module design so that atoms from multiple GROMACS subdomains are aggregated to one rank before the model inference and then perform the deep potential model inference on one rank. Though this simplifies the handling of domain decomposition and ghost atoms, we acknowledge that it may need more invasive domain decomposition schemes in the future for scalability.

Our implementation uses the C++ interfaces of DeePMD-kit and extends the functionalities of GROMACS NNPot module by adding DeePMD-kit inference as a new backend. This implementation supports all DeePMD model families (se\_e2\_a, DPA, DPA2, DPA3 ...) and backends (PyTorch, TensorFlow, JAX ...) in GROMACS MD workflows. Currently, NNPot supports only PyTorch models in TorchScript format on a single MPI rank, and ghost atoms in subdomains are unavailable to the models. Thus, we add a new inference backend to the NNPot module to support users in switching between the original PyTorch backend and our DeePMD backend at compile time. We also modified the NNPot interface to provide ghost-atom information to the inference runtime, which is essential for future parallel inference. These changes ensure long-term maintainability and future compatibility under official commitment.

\section{Evaluation Methodology}
\label{sec:experimentalEnvironment}
We use four MD simulations of four proteins, namely 1YRF, 1UBQ, 3LZM, and 2PTC, to evaluate the \textit{ab initio}-quality enabled by the AI deep potential models integrated in the GROMACS simulation workflow. These proteins offer a wide range in terms of size, folding topology, and biological relevance. Their numbers of atoms are 582, 1231, 2643, and 4114, respectively. They are evaluated in the same GROMACS MD simulation pipeline.

We summarize the key parameters of the simulations in Table~\ref{tab:testcasePara}. The EM stage uses the \texttt{\small steep} integrator and has no fixed number of time steps. Instead, the simulation runs until the system is sufficiently relaxed and the potential energy decreases up to the required threshold. On the other hand, the NVT, NPT, and MD stages use \texttt{\small md} and runs for a user defined number of simulation steps and time. The cutoff range for all short-range interactions during MD is set to 0.7~nm to ensure a safe neighbor‑list margin for the configuration of DPA2 and DPA3 models, which have a cutoff range of 0.6~nm.

In this work, we leverage two latest deep potential models in the GROMACS simulation workflows, i.e., DPA2 (\texttt{\small DPA-2.4-7M}) and DPA3 (\texttt{\small DPA-3.1-3M}), from the DPA model family. DPA2 and DPA3 are foundation models that are pre-trained in a multi-task way, and can be frozen later to branch models specialized in different downstream tasks, as introduced in Section~\ref{sec:Background}. To evaluate their cross-system generalization and transferability, in our MD experiments, we freeze the two DPA models to their \texttt{\small solvated\_protein\_fragments} branch. We then test the \texttt{\small SPICE2} branch derived from DPA2 model, to show the differences in model architectures and branches. Note that \texttt{\small DPA2-\allowbreak solvated\_\allowbreak protein\allowbreak \_fragments} and \texttt{\small DPA2-SPICE2} have the same descriptor but different \texttt{\small fitting-net} from different datasets. In the remainder of this paper, we will use \texttt{\small solvated\_protein} as a short name for \texttt{\small solvated\allowbreak \_protein\allowbreak \_fragments}. 

We evaluate the aforementioned MD simulations on two testbeds using different NVIDIA GPUs. The first testbed is equipped with two A100-40GB GPUs, and we use one GPU in our experiments. This testbed has one AMD EPYC 7302P 16-Core Processor as CPU and 256~GB DDR4 main memory. The second testbed is equipped with a NVIDIA GH200 480~GB superchip. This GH200 has a 72 core ARM64-based Grace CPU with 480~GB LPDDR5X memory and a Hopper 100 GPU (H100) with 96~GB HBM3 memory. CPU and GPU are interconnected via cache-coherent NVLink-C2C to provide 900 GB/s bandwidth, which also provides the potential for model inference utilizing the CPU-GPU unified-memory \cite{10.1145/3673038.3673110}. Since most of the ML inference workload is performed on the GPU, this setup makes them well-suited for demonstrating the performance of deep-potential AIMD simulations workflows on today's GPU-based HPC systems.

We summarize the software environment information in Table~\ref{tab:softENV}. GROMACS is compiled with CUDA GPU support. All MD runs use a single MPI process and 8 OpenMP threads. PyTorch was used as the backend of DeePMD-kit for all the model inference stage, as it is the native backend of both DPA2 and DPA3 models.

\begin{table}[htb]
\caption{Specification of Experiments}
\begin{subtable}[t]{.245\textwidth}
\caption{Test cases parameters\label{tab:testcasePara}}
\raggedleft
\resizebox{1\linewidth}{!}{
\begin{tabular}{@{\extracolsep{1pt}}lcccc}
    \toprule
        & EM  & NVT \& NPT & MD \\
    \midrule
        $\Delta$ t (fs)  & -  & 2 & 1 \\
        Steps & - & 50K & 10K \\
        Total Time (ps) & -  & 100 &  10 \\
        rcutoff (nm) &  1.2  & 1.2  & 0.7 \\
        NN Pot & No  & No  & Yes  \\
        NN Atom Group & - & - & Protein \\
    \bottomrule
    \end{tabular}
    }
\end{subtable}
\hfill
\begin{subtable}[t]{.235\textwidth}
\caption{Software Environments\label{tab:softENV}}
\raggedright
\resizebox{\linewidth}{!}{
    \begin{tabular}{@{\extracolsep{1pt}}lcccc}
    \toprule
        & A100 & GH200 \\
    \midrule
       GROMACS  & 2025.2 & 2025.2 \\
       DeePMD-kit  & v3.1.0 & v3.1.0 \\
       PyTorch  & 2.7.1 & 2.5.1\\
       CUDA  & 12.6 & 12.4\\
       GCC  & 12.2 & 11.4\\
       OS/Linux & CentOS8/4.18 & RHEL9.4/5.14 \\
    \bottomrule
    \end{tabular}
}
\end{subtable}
\end{table}

We use FP32 as the main data precision for GROMACS and DeePMD parts in all tests. For each protein benchmark, we run multiple tests on a combination of proteins, models, and hardware testbeds. To provide a reference of MD simulation without quantum-mechanical effects, we also perform classical MD simulations that only use empirical fitting to quantum laws for computing potential and pseudo-potential. Some DPA3 based tests cannot be run on particular hardware due to their GPU memory requirements exceeding the available memory, we provide additional details in Section~\ref{sec:results}. 

Simulation performance is measured using GROMACS’s built-in timing and performance modules, which report the ratio of simulated time to actual wall-clock time in units of ns/day. This metric, commonly used in molecular dynamics benchmarking, offers a clear and consistent measure of simulation throughput, where higher values indicate better performance. 

GPU memory usage is traced at a frequency of every 10 seconds using the \texttt{\small nvidia-smi} command-line tool. The reported memory usage reflects the combined consumption of both GROMACS MD run and the DeePMD model inference. To provide in-depth characterization of deep potential models in the MD workflow, we use Nsight Systems and Nsight Compute to profile the first 200 steps of the MD simulations. Additionally, we employ NVTX annotations to mark model inference calls within the DeePMD-kit C++ source code. These annotations are then captured and visualized to enable precise identification of the corresponding simulation steps.

\section{Evaluation Results}
\label{sec:results}

\begin{figure*}[h]
    \centering
    \includegraphics[width=\textwidth]{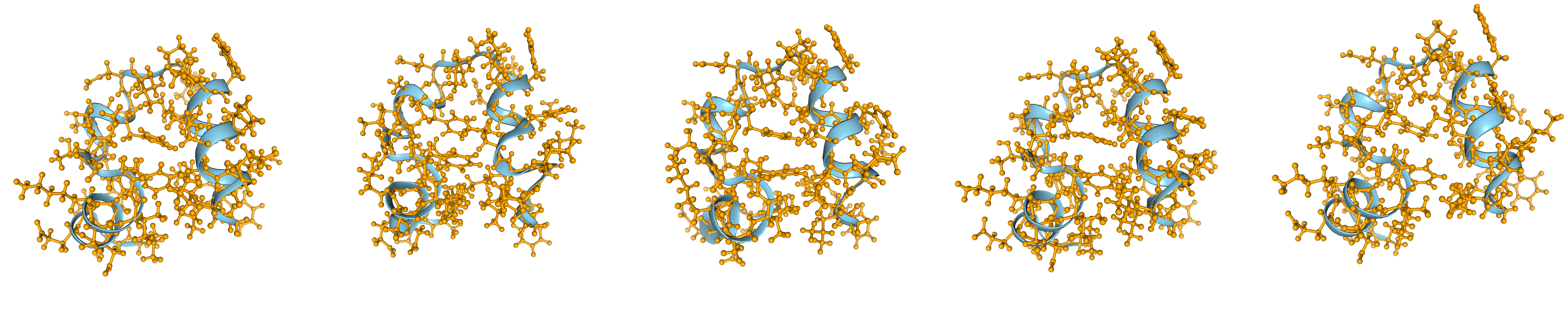}
    \caption{Snapshots of 1YRF protein after 10 ps MD simulations. The initial structure (leftmost) is compared against trajectories from DPA2-solvated\_protein, DPA2-SPICE2 , DPA3-solvated\_protein, and the empirical GROMACS force field (rightmost). Note that deviations in local atomic positions between DP models and the empirical field are expected, as DP models incorporate ab initio-level quantum effects.}
    \label{fig:1yrf}
\end{figure*}

In this section, we first validate the AIMD simulation workflow and then evaluate the main ML MD inference stage in terms of simulation throughput and scalability. We further provide a performance characterization across two GPU platforms, including kernel execution and memory usage for the DPA2 and DPA3 models.

We evaluate the correctness of the \textit{ab initio}-quality simulation workflow in GROMACS by comparing the DPA-based quantum mechanical force fields with empirically simulated force fields in the four protein benchmarks. Figure~\ref{fig:1yrf} visualizes solvated protein 1YRF after a 10~ps MD run with quantum mechanical effects enabled by DPA2 and DPA3, respectively, and compared to empirically simulated results, where the orange spheres are atoms with connecting bonds, and blue ribbons represent the protein's secondary structure. From the three snapshots, we can verify that the $\alpha$-helical secondary structures (blue ribbons) near the center and lower portion of the protein are consistent across the four simulation setups. In addition, DPA2-SPICE2 and DPA2-solvated\_protein are two downstream models obtained via freezing a shared representation, which is obtained from training on diverse and general-purpose datasets. By examining these two downstream models, we validate the fundamental properties of LAMs, such as cross-system generalization and transferability.

\subsection{Comparison of Deep Potentials Models}
\label{sec:perf}
\begin{figure}[bt]
    \centering
    \includegraphics[width=1\linewidth,height=0.7\linewidth]{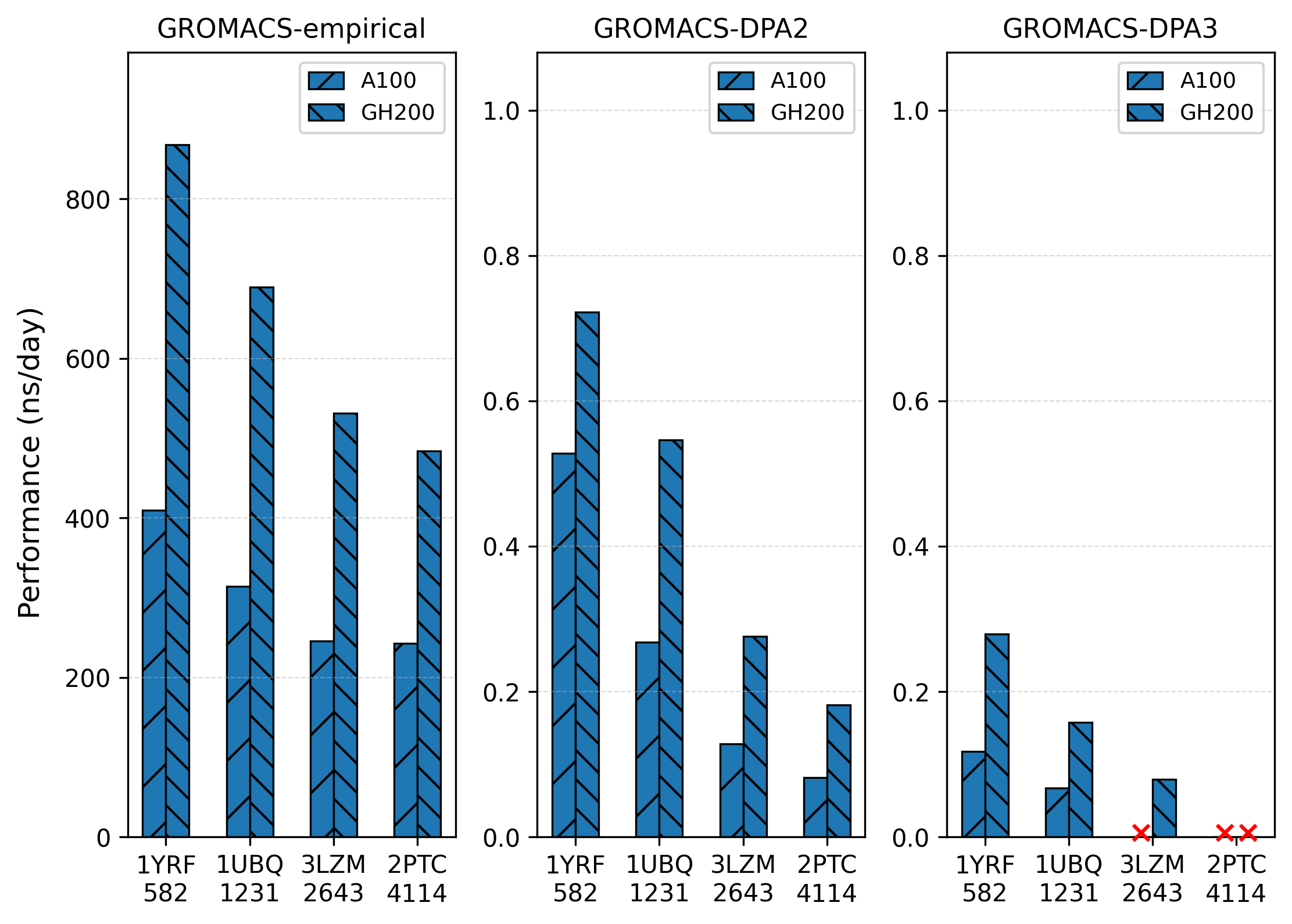}
    \caption{MD simulation throughput for classical approach, \textit{ab initio} with DPA2, and \textit{ab initio} with DPA3 for four protein simulations on A100 and GH200. DPA3 results marked with $\times$ are unavailable due to OOM.}
    \label{fig:speed_compare}
\end{figure}

We evaluate the simulation throughput of the four protein benchmarks using two LAMs, DPA2 and DPA3, on A100 and GH200 GPU architectures. Figure~\ref{fig:speed_compare} presents the MD simulation throughput in ns/day on two GPU testbeds. The highest throughput, 0.7 ns/day is achieved by the DPA2 model on GH200 in the case of the 1YRF protein. As the protein benchmarks increase the number of atoms from 582 atoms, to eventually 4,114 atoms, the simulation throughput decreases to 0.2 ns/day. On A100 GPU, the DPA2 model achieves a simulation throughput that changes from 0.55 ns/day for the 1YRF protein to 0.1 ns/day for the largest protein 2PTC. Note that since \texttt{\small DPA2-\allowbreak solvated\_protein} and \texttt{\small DPA2-SPICE2} are branches derived from the same DPA2 model, they share the same neural network architecture and therefore exhibit exactly the same throughput. On both GPU testbeds, the DPA2 models exhibit an approximately linear trend of scaling simulation throughput as the number of atoms increases. This linear scaling is consistent with the theoretical expectation of ML inference in DPA2 models. Compared to the the cubic scaling of first-principle quantum-mechanical methods like density functional theory (DFT), the simulation throughput obtained from DPA models confirms the superior scaling properties of deep potential models in \textit{ab initio}-quality MD simulation. 

Compared to DPA2, DPA3 model based on a graph neural network (GNN), consistently achieves lower throughput than DPA2 across all protein benchmarks. In particular, DPA2 outperforms DPA3 by an average factor of $4.23\times$ on A100 and $3.18\times$ on GH200. Similar to DPA2, the simulation throughput by DPA3 also decreases as the size of the tested protein increases and exhibit an approximately linear scaling. Unlike DPA2, DPA3 cannot finish simulating the 3LZM and 2PTC proteins on A100, neither the 2PTC protein on GH200 due to out-of-memory (OOM) errors on GPU. We present the characterization of their GPU memory usage in Section~\ref{sec:memory}.

DPA2 and DPA3 based MD simulations exhibit different speedup when moving from a small GPU architecture to a larger GPU architecture. Despite the peak GPU throughput improves from 19.5 FP32 TFLOPS to 60 TFLOPS across these two testbeds, for the 1YRF protein benchmark, the simulation throughput by DPA2 merely increases from 0.55~ns/day to 0.7~ns/day. However, other protein benchmarks by DPA2 exhibit higher speedup when moving from A100 to GH200, i.e., $2.2\times$ in 1UBQ , $2.9\times$ in 3LZM and $2\times$ in 2PTC. Since the other benchmarks have $2\times$-$7\times$ 1YRF's atoms, and thus higher computational load, they can better utilize the increased GPU capacity on GH200. Unlike DPA2, even for the 1YRF protein, moving from A100 to GH200 increases the simulation throughput by almost $3\times$. A DPA3 is based on graph neural network architecture, the improved memory bandwidth of HBM3 on GH200 likely contributes more substantially to the performance improvement.   

To understand the computational overhead needed for enabling \textit{ab initio} quality, we also configure a classical MD simulation and measure its throughput on two GPU testbeds. The classical MD achieves a throughput between 200 and 400 ns/day for the A100 testbed, and 500 to 900 ns/day for the GH200 testbed. The substantial performance reduction from introducing deep potentials indicate that although ML inference can reduce the high computational cost from $\mathcal{O}{(N^3)}$ in DFT, the AIMD models are still computationally more expensive compared to empirical MD simulations. 

\observation{\textit{Deep potentials based AIMD simulations exhibit a throughput scaling approximately linearly with the number of atoms, unlike the cubic scaling of DFT.}}

\observation{\textit{The DPA2 deep potential model consistently achieves $3–4\times$ higher simulation throughput than DPA3 across all protein benchmarks.}}

\subsection{Characterization of Deep Potential MD on GPUs}
\begin{table}[bt]
  \centering
  \caption{A breakdown of top GPU kernels from profiling 200-step MD of 1UBQ using DPA2 and DPA3 on A100.}
  \label{tab:kernel_summary}
  \setlength{\tabcolsep}{2pt} 
  \renewcommand{\arraystretch}{0.9} 
  \small
  \begin{tabular}{l r r r r}
    \toprule
    \multirow{2}{*}{\textbf{Kernel Type}} & \multicolumn{2}{c}{\textbf{DPA2}} & \multicolumn{2}{c}{\textbf{DPA3}} \\
    \cmidrule(lr){2-3} \cmidrule(lr){4-5}
     & Time (\%) & Launches & Time (\%) & Launches \\
    \midrule
    GEMM & 44.1 & 140k & 8.0 & 210k \\
    Element-wise \& Indexing & 40.9 & 600k & 75.5 & 1.7M \\
    Reduce  & 7.5 & 120k & 10.0 & 280k  \\
    Other & 7.5 & 140k & 6.5 & 410k \\
    \midrule
    \textbf{Total} & \textbf{100} & \textbf{1.0M} & \textbf{100} & \textbf{2.6M} \\
    \bottomrule
  \end{tabular}
\end{table}

We profile the kernel execution patterns on GPUs to understand their performance characteristics in DPA2 and DPA3. Profiling traces for DPA2 and DPA3 indicate that the MD simulation runtime is mostly dominated by deep potential model inference. For the DPA2 model, GPU kernel execution accounts for 96\% of the total runtime, compared to 98.4\% for DPA3. Table~\ref{tab:kernel_summary} presents a breakdown of kernel execution time and kernel launches in the two models using a MD simulation of the 1UBQ protein. 

In DPA3, element-wise and indexing kernels dominate over 75\% runtime. Indexing kernels, such as \texttt{\small indexFuncLargeIndex} and \texttt{\small index\allowbreak SelectLargeIndex}, alone account for over 34\% of the kernel execution time. Fused and scatter/gather operations also contribute a large share to the runtime. These kernels are primarily responsible for rearranging, selecting, scattering, or gathering data in memory. Therefore, they underutilize the computing power of GPUs but stress more the memory subsystem. GEMM related computational kernels are also present in DPA3, but only account for about 8\% runtime. This kernel characteristics in DPA3 is consistent with its graph neural network layers the architecture. Thus, DPA3’s memory bound operations potentially benefit from hardware with high memory bandwidth. DPA2 workflows show a balanced mix of dense matrix multiplications and indexing operations. It spends $44.1\%$ runtime in GEMM kernels, with element-wise and indexing kernels taking $40.9\%$ runtime. The GEMM-heavy nature suggest that DPA2 could potentially benefit from hardware with tensor-core acceleration. Both DPA workflows have a substantially higher number of kernel launches, i.e., 2.61\,M in DPA3 and \ 1.01\,M in DPA2, indicating increased launch overhead and reduced kernel fusion efficiency. A typical optimization technique for high launch overhead is to leverage CUDA Graphs. However, profiling results indicate no CUDA Graphs is utilized in the current backend engine.

The kernel characteristics highlights different optimization priorities in the two models. For instance, kernel fusion and GEMM tuning are the main optimization opportunities in DPA2. DPA3 could explore optimizations via CUDA Graphs to reduce kernel launch overhead and fusion of small element-wise kernels to improve the overall GPU occupancy. Given the repetitive and predictable nature of deep potential models, these optimizations may yield substantial performance gains. One limitation is the lack of support for domain decomposition for deep potential models, especially DPA family. Therefore, all atoms need to be processed on one GPU for inference and thus limiting scalability and hardware portability. Future work should focus on enabling distributed inference with ghost atoms across MPI ranks. 

\observation{\textit{DPA2 workflow is balanced among GEMM-heavy and indexing operations while DPA3 workflow is dominated by memory-bound indexing operations.}}

\subsection{GPU Memory Usage}
\label{sec:memory}
\begin{figure}[bt]
    \centering
    \includegraphics[width=0.85\linewidth]{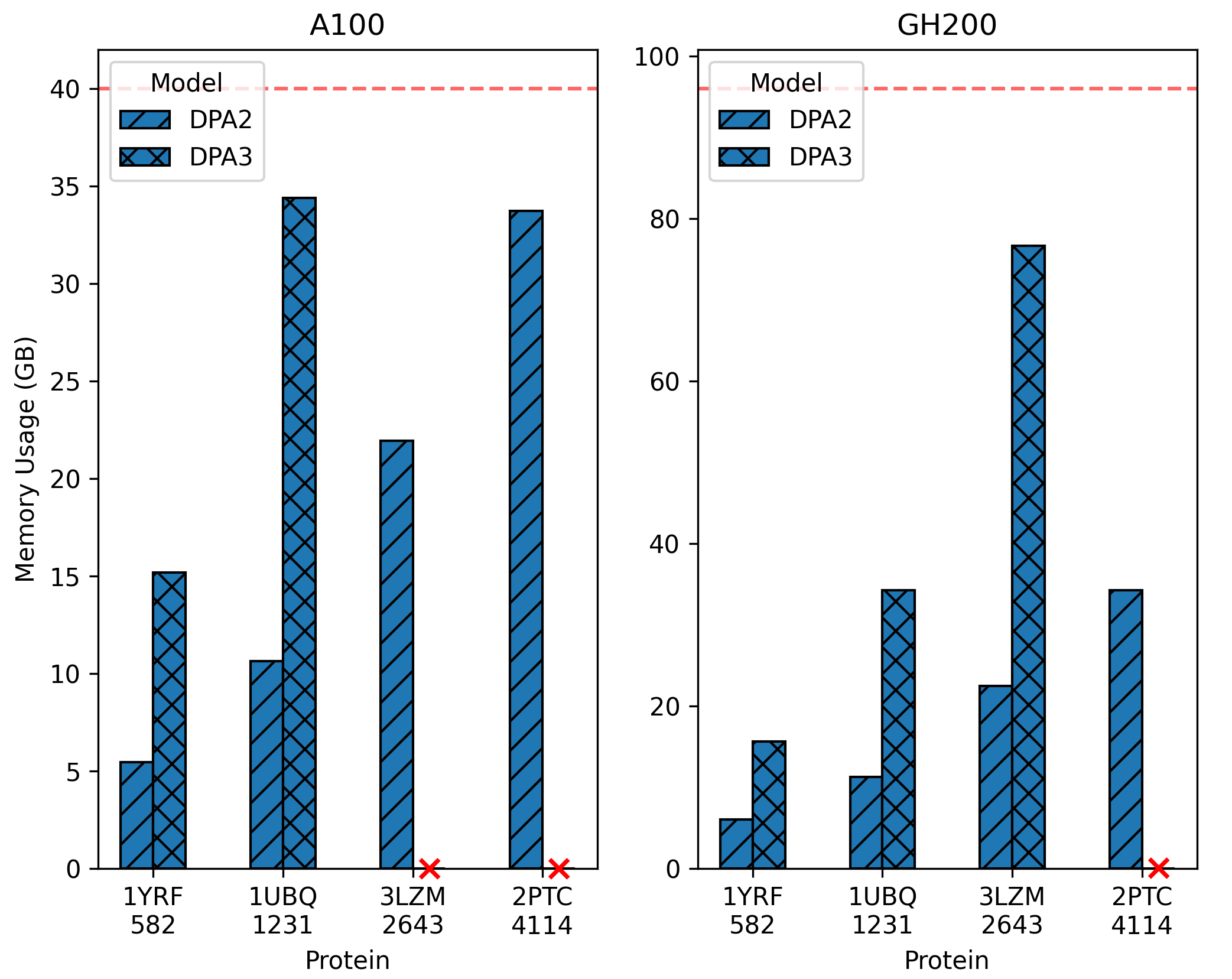}
    \caption{GPU memory usage of DPA2 and DPA3 models on A100 and GH200. The red dashed line represents the GPU capacity. Results marked with $\times$ are unavailable due to OOM.}
    \label{fig:vram}
\end{figure}

From the kernel characterization, we find DPA2 and DPA3 both have substantial memory-bound indexing operations. Together with DPA3's frequent OOM errors (Section~\ref{sec:perf}), we decide to further investigate the impact of GPU memory on the DPA MD workflows. Figure~\ref{fig:vram} presents the GPU memory usage in DPA2 and DPA3 on the A100 and GH200 testbeds. The GPU memory usage of both DPA MD workflows are high. For the smallest protein benchmark of 582 atoms, the DPA2 workflow requires approximately 5~GB GPU memory. For the other larger benchmarks, the GPU memory usage increases to 10, 22, and eventually 34~GB on A100 GPU. The memory usage of DPA2 remain the same across the two GPU testbed. 

DPA3 consistently uses significantly more memory than DPA2, reaching about $3\times$ memory usage for the same protein simulation. For instance, it uses 15~GB memory for the smallest protein benchmark, 35~GB memory for 1UBQ protein and 70~GB memory for 3LZM protein. Due to this higher memory footprint, the 3LZM protein with the DPA3 model can only be run on the GH200 system. For the largest structure, the 2PTC protein, the memory requirement of DPA3 is estimated to be more than 140~GB and thus exceeds the 96~GB capacity on GH200, preventing the experiment from being executed.

The memory demand for both DPA models scales linearly with the number of atoms. This is as expected because both model descriptors use message-passing neural network. In the core iterations of these two models, the representations of neighbor atoms are used by each other as input for the next update. It means the representation of an atom $i$ can be affected by not only its neighbor $j$ but also the neighbor atoms of $j$, as the information is encoded into the representation of $j$ in previous updates. In other words, the receptive field of an atom can be expanded beyond its neighbor area, to capture longer relationships in the system, while it might cause negative effects in computations and scalability.

The high GPU memory usage will be the primary limiting factor in production deep potential AIMD simulations. This highlights the necessity for memory-aware model design or the parallel execution strategies for larger systems. This finding indicates that large-scale simulations will only be feasible with either distributed memory systems or heterogeneous memory system that exploit large CPU side memory, such as NVIDIA GH200 \cite{10.1145/3673038.3673110} and AMD MI300A \cite{11242068}. Future optimizations such as mixed-precision inference or activation check-pointing could further alleviate these memory constraints.

\observation{\textit{DPA2 and DPA3 exhibit linear memory usage scaling with the number of atoms. DPA3 needs $3\times$ more GPU memory than DPA2, limiting its feasibility for large simulations.}}

\section{Related Work}
 
MD software such as GROMACS~\cite{pall2020gromacs} and LAMMPS~\cite{thompson2022lammps} exploit multi-level parallelism across CPU and GPU on HPC systems for performance. 
The emergence of deep neural network potentials including DeePMD model families has enabled near-DFT accuracy at an relatively low cost for MD simulations in multiple fields \cite{zhang_deep_2018, chang_efficient_2024}. Several MD software now provides support deep potentials, either via custom interfaces or through the coupling with DeePMD-kit and similar frameworks. For example, DeePMD-kit has provided integration to multiple MD software packages including LAMMPS, i-PI and AMBER ~\cite{zeng_deepmd-kit2_2023}, while many of them are developing their own interfaces for deep potentials. To the best of our knowledge, this work presents the first integration of the latest GROMACS and DeePMD-kit through the official GROMACS interface, as well as the first comprehensive GPU-level performance characterization of the DPA2 and DPA3 models in protein molecular dynamics simulations. 

Several studies have explored the performance and scalability of deep potential models on various HPC systems. Jia \textit{et al.}~\cite{weile2020deepMD_lammps_gordonbell_100milion} demonstrate large-scale molecular dynamics simulations with deep potentials reaching 100 million atoms on NVIDIA V100-accelerated systems. Du \textit{et al.}~\cite{du_scaling_2025} enable 500 million atoms on a 4096-node ARMv8 cluster, leveraging the ARM SVE vectorization instruction set to accelerate deep potential inference. 
However, most of the studies focused on earlier deep potential models, leaving the emerging attention-based and graph-neural-network (GNN) DPA model families unexplored. Furthermore, the majority of benchmark systems examined so far are limited to materials such as water and copper, while the performance of deep potentials in biomolecular simulations remains unexplored. This work fills these gaps by investigating attention-based DPA2 and GNN based DPA3 models on protein systems.

\section{Conclusion}
\label{sec:discussionConclusions}
In this work, we enabled AI deep potentials in GROMACS by coupling the DeePMD-kit framework with GROMACS neural network potentials. We first validated the workflow on GPUs on multiple protein systems, showing correct force evaluations and stable MD trajectories. We further evaluated two recent deep potential large-atom models, DPA2 and DPA3, on four solvated protein MD simulations using the proposed AIMD workflow to enable \textit{ab initio} quality. Our results show that DPA2 delivers up to $4.23\times$ and $3.18\times$ higher throughput than DPA3 on A100 and GH200 GPUs. We provided a first-time characterization of end-to-end GROMACS MD workflows on GPU with deep potentials {ab initio} quality. Our results evaluated the fundamental properties of LAMs, such as cross-system generalization and transferability, by employing two downstream models derived from the same pre-trained DPA-2 foundation model. Finally, we identify kernel-launch overhead and domain-decomposed inference as the main optimization priorities for AI deep potentials in MD workflows.

\section*{Acknowledgment}
This research is supported by SeRC (Swedish eScience Research Centre) SESSI Efficient Simulation Software Initiative.

\bibliographystyle{IEEEtran}
\bibliography{bibliography.bib}

\end{document}